# Complete-return spectrum for a generalized Rosen-Zener two-state term-crossing model

## T.A. Shahverdyan[1], D.S. Mogilevtsev[2], V.M. Red'kov[2], and A.M Ishkhanyan[3]

[1]Moscow Institute of Physics and Technology, 141700 Dolgoprudni, Russia
[2]Institute of Physics, NASB, F. Skarina Avenue 68, Minsk 220072, Belarus
[3]Institute for Physical Research of NAS of Armenia, 0203 Ashtarak, Armenia

The general semiclassical time-dependent two-state problem is considered for a specific field configuration referred to as the generalized Rosen-Zener model. This is a rich family of pulse amplitude- and phase-modulation functions describing both non-crossing and term-crossing models with one or two crossing points. The model includes the original constant-detuning non-crossing Rosen-Zener model as a particular case. We show that the governing system of equations is reduced to a confluent Heun equation. When inspecting the conditions for returning the system to the initial state at the end of the interaction with the field, we reformulate the problem as an eigenvalue problem for the peak Rabi frequency and apply the Rayleigh-Schrödinger perturbation theory. Further, we develop a generalized approach for finding the higher-order approximations, which is applicable for the whole variation region of the involved input parameters of the system. We examine the general surface $U_{0n} = U_{0n}(\delta_0, \delta_1)$, $n = \mathrm{const}$, in the 3D space of input parameters, which defines the position of the $n$-th order return-resonance, and show that for fixed $\delta_0$ the curve in $\{U_{0n}, \delta_1\}$ plane, i.e., the $\delta_0 = \mathrm{const}$ section of the general surface is accurately approximated by an ellipse which crosses the $U_{0n}$-axis at the points $\pm n$ and $\delta_1$-axis at the points $\delta_{11}$ and $\delta_{12}$. We find a highly accurate analytic description of the functions $\delta_{11}(\delta_0, n)$ and $\delta_{12}(\delta_0, n)$ as the zeros of a Kummer confluent hypergeometric function. From the point of view of the generality, the analytical description of mentioned curve for the whole variation range of all involved parameters is the main result of the present paper.

The general semiclassical time-dependent two-state problem [1,2] is written as a system of two coupled first-order differential equations for probability amplitudes $a_1(t)$ and $a_2(t)$ of the first and second states, respectively, containing two arbitrary real functions of time, $U(t)$ (amplitude modulation; $U > 0$) and $\delta(t)$ (phase modulation):

$$i\frac{da_1}{dt} = Ue^{-i\delta}a_2, \quad i\frac{da_2}{dt} = Ue^{+i\delta}a_1. \tag{1}$$

When discussing the excitation of an effective two-state quantum system by an optical laser field, the amplitude modulation function $U(t)$ is referred to as the Rabi frequency and the derivative $\delta_t(t) = d\delta / dt$ of the phase modulation is the detuning of the laser frequency from the transition frequency. Below, we discuss the following field configuration

$$U(t) = U_0 \mathrm{sech}(t), \quad \delta_t(t) = \delta_0 + \delta_1(\mathrm{sech}(t))^2 \tag{2}$$

with constant $U_0 > 0$, $\delta_{0,1}$. We refer to this model as the generalized Rosen-Zener model because it includes the original constant-detuning Rosen-Zener model [3] as a particular case when $\delta_1 = 0$. The time-variation shape of the detuning for several values of the parameter $\delta_1$ is shown in Fig.1. As it is seen, this is a rich family describing both non-crossing and term-crossing models with one or two crossing points. The detuning does not cross zero if $-\delta_0/\delta_1 < 0$ or $-\delta_0/\delta_1 > 1$, while at $\delta_0 + \delta_1 = 0$ it touches the origin and there exist two crossing points located at $t = \pm \mathrm{arcsech}[\sqrt{-\delta_0/\delta_1}]$ if $0 \le -\delta_0/\delta_1 < 1$. Finally, we note that the particular case $\delta_0 = 0$ is an exceptional one since then the detuning asymptotically goes to zero as $t \to \mp\infty$.

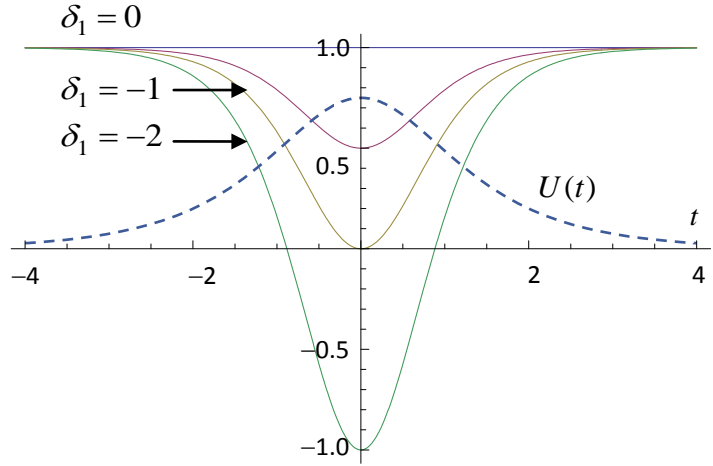

Fig.1. Generalized Rosen-Zener model defined by Eq. (2): $U_0 = 0.75$, $\delta_0 = 1$.

System (1) is equivalent to the following linear second-order ordinary differential equation

$$\frac{d^2 a_1}{dt^2} + \left( i\frac{d\delta}{dt} - \frac{1}{U}\frac{dU}{dt} \right)\frac{da_1}{dt} + U^2 a_1 = 0. \qquad (3)$$

The transformation of the independent variable $z = (1 + \tanh(t))/2$ reduces this equation to the confluent Heun equation [4]

$$\frac{d^2 a_1}{dz^2} + \left( 2i\delta_1 + \frac{(1+i\delta_0)/2}{z} + \frac{(1-i\delta_0)/2}{z-1} \right)\frac{da_1}{dz} - \frac{U_0^2}{z(z-1)}a_1 = 0. \qquad (4)$$

If $\delta_1 = 0$, this equation is reduced to the Gauss hypergeometric equation. Accordingly, the solution satisfying the initial conditions $a_1(t=-\infty) = 1$, $a_1'(t=-\infty) = 0$ discussed here is written as

$$a_{1RZ} = {}_2F_1(-U_0, U_0; (1 + i\delta_0)/2; z) \ , \tag{5}$$

from which the known formula by Rosen and Zener [3] for the final transition probability to the second state $|a_2(t)|^2$ is derived:

$$P_{RZ} = (\sin(\pi U_0)\text{sech}(\pi\delta_0/2))^2 \ . \tag{6}$$

This formula states the known $\pi$- theorem [1,2] according to which the system returns to the initial state $(a_1, a_2) = (1,0)$ if $U_0 = n$, $n = 0,1,2... \in N$. Note that if the population inversion (i.e., a total transition to the second state) is discussed, the situation is different. The peculiarity here is that the system is completely inverted only at exact resonance, $\delta_0 = 0$. Furthermore, it is seen that the maximal transition to the second state allowed for the given fixed value of the detuning is achieved at $U_0 = 1/2 + n$. It should be noted that the return-to-the-initial-state theorem comes already from the discussion of the Rabi box-model with constant amplitude and detuning modulations [5]. Furthermore, the solution of the problem for the exact resonance case reveals that for any pulse of duration $T$ the transition probability is given as $p \sim (\sin(A))^2$, where $A = \int_0^T U(t)dt$, i.e., $A$ is the area under the pulse envelope.

This formula shows that, indeed, the transition probability vanishes when $A$ is a multiple of $\pi$ and for half-integer multiples of $\pi$ we get a complete population inversion. It was conjectured that this is a general property. However, the close inspection has revealed that in general this is not the case. Both the statement concerning the inversion and the one regarding the return are violated for asymmetric pulses while for symmetric pulses the return property survives with a changed value of the pulse area (see, e.g., the discussion of this point in [6]). Qualitatively the same picture is observed when a quadratic-nonlinear extension of the two-state approximation is discussed in the framework of two-mode photo- or magnetic-association of ultracold atoms in degenerate quantum ensembles [7].

When inspecting the conditions for returning to the initial state at the end of the interaction with the field, we note that the additional boundary condition $|a_1(t = +\infty)| = 1$ together with the initial conditions defines an eigenvalue problem for the peak Rabi frequency $U_0$ that can be written as

$$\hat{H}_0 a_1 - U_0^2 a_1 = -2i\delta_1 z(z-1)\frac{da_1}{dz} \ , \tag{7}$$

$$a_1(t = -\infty) = 1, \quad a_1'(t = -\infty) = 0, \quad |a_1(+\infty)| = 1, \tag{8}$$

where the operator $\hat{H}_0$ stands for the standard constant-frequency Rosen-Zener model:

$$\hat{H}_0 = z(z-1)\left[\frac{d^2}{dz^2} + \left(\frac{(1+i\delta_0)/2}{z} + \frac{(1-i\delta_0)/2}{z-1}\right)\frac{d}{dz}\right]. \tag{9}$$

Having at hand the solution of the problem for the Rosen-Zener case with the eigenfunctions of the problem given by equation (5) and eigenvalues $U_0 = n$ one may apply the Rayleigh-Schrödinger perturbation theory to examine, approximately, the behavior of the system for non-zero $\delta_1$, at least, for small $\delta_1$. Thus, we suppose that $\delta_1$ is small enough and for the $n$ th-order complete return case apply the expansion

$$a_{1n} = a_{1RZn} + (-2i\delta_1)u_1 + (-2i\delta_1)^2 u_2 + \dots, \tag{10}$$

$$U_{0n}^2 = U_{0RZn}^2 + (-2i\delta_1)U_1 + (-2i\delta_1)^2 U_2 + \dots, \tag{11}$$

where, according to Eq. (5),

$$a_{1RZn} = {}_2F_1(-n, n; (1+i\delta_0)/2; z) \quad \text{and} \quad U_{0RZn} = n. \tag{12}$$

Equation (7) then reads

$$\left(\hat{H}_0 - (n^2 + (-2i\delta_1)U_1 + (-2i\delta_1)^2 U_2 + ..)\right)(a_{1RZn} + (-2i\delta_1)u_1 + (-2i\delta_1)^2 u_2 + ...)$$
$$= -2i\delta_1 z(z-1)\frac{d}{dz}(a_{1RZn} + (-2i\delta_1)u_1 + (-2i\delta_1)^2 u_2 + ...) \tag{13}$$

so that equating the terms at equal powers of $\delta_1$ we obtain

$$\left(\hat{H}_0 - n^2\right)a_{1RZn} = 0, \tag{14}$$

$$\left(\hat{H}_0 - n^2\right)u_1 - U_1 a_{1RZn} = z(z-1)\frac{da_{1RZn}}{dz}, \tag{15}$$

$$\left(\hat{H}_0 - n^2\right)u_2 - U_1 u_1 - U_2 a_{1RZn} = z(z-1)\frac{du_1}{dz} \tag{16}$$

and so on for higher orders. Eq. (14) is, of course, automatically satisfied for any $n$. Now, to treat Eq. (15) for $u_1$, we apply an expansion in terms of functions $a_{1RZn}$:

$$u_1 = \sum_{m=0}^{\infty} b_m a_{1RZm}, \quad b_m = \text{const}. \tag{17}$$

Note now that the functions $a_{1RZn}$ are orthogonal in the interval $z \in [0,1]$ with the weight function $w(z) = z^{-(1-i\delta_0)/2}(z-1)^{-(1+i\delta_0)/2}$:

$$\int_0^1 w(z)a_{1RZm}a_{1RZn}\,dz = C_n\delta_{mn}, \tag{18}$$

where $\delta_{mn}$ is the Kronecker delta and $C_n$ is a constant. Substituting then expansion (17) into Eq. (15), multiplying the resultant equation by $w(z)a_{1RZm}$ and integrating over the interval $z \in [0,1]$ we obtain

$$(m^2 - n^2)b_m C_m = U_1 C_n \delta_{mn} + \int_0^1 z(z-1)w(z)\frac{da_{1RZn}}{dz}a_{1RZm}\,dz\,. \qquad (19)$$

Taking here $m = n$ we get the first correction for the Rabi frequency:

$$U_1 = -\frac{1}{C_n}\int_0^1 z^{(1+i\delta_0)/2}(z-1)^{(1-i\delta_0)/2}\frac{da_{1RZn}}{dz}a_{1RZn}\,dz\,, \qquad (20)$$

and for $m \neq n$ we determine the coefficients of expansion (17):

$$b_m = \frac{V_{mn}}{(m^2 - n^2)C_m}\,, \quad V_{mn} = \int_0^1 z(z-1)w(z)\frac{da_{1RZn}}{dz}a_{1RZm}\,dz\,. \qquad (21)$$

In the first approximation, according to Eq. (11), the Rabi frequency supporting the return of the system to the initial state is given as $U_{0n} = \sqrt{n^2 - 2i\delta_1 U_1}$. Calculation of the integrals involved in Eq. (20) leads to the following simple final result

$$U_{0n} = n\sqrt{1 - \frac{2\delta_0\delta_1}{4n^2 - 1}}\,. \qquad (22)$$

Numerical simulations show that this is a satisfactory approximation as long as the parameter $\delta_1$ is small compared with $U_0^2 = n^2$. However, at large $\delta_0$ and $\delta_1$, one needs a more accurate result, especially, for lower-order return resonances. For this reason, we calculate the next correction term $U_2$ using Eq. (16). Again, applying for $u_2$ the expansion

$$u_2 = \sum_{m=1}^{\infty} d_m a_{1RZm}\,, \quad d_m = \text{const} \qquad (23)$$

and multiplying the equation by $w(z)a_{1RZm}$ with $m = n$ and integrating we obtain

$$-U_1 b_n C_n - U_2 C_n = \sum_{m=0}^{\infty} b_m V_{mn}\,. \qquad (24)$$

Since $-U_1 C_n = V_{nn}$, for the second-order correction to the Rabi frequency we obtain

$$U_2 = -\frac{1}{C_n}\sum_{m=0,m\neq n}^{\infty}\frac{V_{mn}^2}{(m^2 - n^2)C_m}\,. \qquad (25)$$

Unfortunately, the integrals involved in this sum are not calculated analytically. However, we will now show that the final sum itself can be calculated with remarkably high accuracy using a different approach. The crucial observation here is that the result of the second-order approximation can be rewritten as a bilinear form with respect to $U_{0n}$ and $\delta_1$:

$$\frac{U_{0n}^2}{n^2} = \left(1 - \frac{\delta_1}{\delta_{11}}\right)\left(1 - \frac{\delta_1}{\delta_{12}}\right), \tag{26}$$

where the functions $\delta_{11} = \delta_{11}(\delta_0, n)$ and $\delta_{12} = \delta_{12}(\delta_0, n)$ are the roots of the equation $n^2 + (-2i\delta_1)U_1 + (-2i\delta_1)^2 U_2 = 0$. For a fixed $\delta_0$ Eq. (26) defines a second-degree curve in the $\{U_{0n}, \delta_1\}$-plane, which in our case may be an ellipse or parabola. In order to find the functions $\delta_{11}$ and $\delta_{12}$ which define the actual form of the curve, we explore the limit $U_0 \to 0$ of the problem for vanishing but *non-zero* $U_0$. Note that the axis $U_{0n} \equiv 0$ corresponds to the non-physical case of an interaction without a laser field. Mathematically, it is seen that the axis $U_{0n} = 0$ presents a degenerate case when each point of the axis satisfies the return condition $|a_1(+\infty)| = 1$ because the solution of the problem in this case is $a_1(t) \equiv 1$. It is for this reason that one should explore the vanishing limit $U_0 \to 0$ for non-zero $U_0$.

Considering the last term in Eq. (4) as a perturbation and applying the method of variation of constants, we find that the solution of the problem in the limit $U_0 \to 0$ satisfying the applied initial conditions in the first approximation is written as

$$a_1 = 1 - y_1 \int_0^z \frac{F(z) y_2}{W} dz + y_2 \int_0^z \frac{F(z) y_1}{W} dz, \tag{27}$$

where $F(z) = U_0^2 / (z(z-1))$ and $W$ is the Wronskian of fundamental solutions

$$y_1 = 1, \quad y_2 = \int_0^z e^{-2i\delta_1 z} z^{-(1+i\delta_0)/2} (z-1)^{-(1-i\delta_0)/2} dz \tag{28}$$

of the homogeneous equation, i.e., Eq. (4) when $U_0$ is put equal to zero. Note that $W = d y_2 / d z$. The first integral involved in Eq. (27) is symmetric with respect to the change $z \to 1-z$, hence it vanishes in the limit $z \to 1$ ($t \to +\infty$). Further, it can be shown that for $z = 1$ the last term in Eq. (27) is written in terms of the Kummer confluent hypergeometric functions. The final result reads

$$a_1(t = +\infty) = 1 - \pi^2 U_0^2 (\operatorname{sech}(\pi \delta_0 / 2))^2 |_1 F_1((1 + i\delta_0)/2; 1; 2i\delta_1)|^2. \tag{29}$$

From here, we conclude that in the limit of infinitesimally small but non-zero $U_0$ complete return occurs if

$$_1F_1((1+i\delta_0)/2;1;2i\delta_1) = 0 \,. \tag{30}$$

Hence, the functions $\delta_{11} = \delta_{11}(\delta_0, n)$ and $\delta_{12} = \delta_{12}(\delta_0, n)$ are the roots of this equation. Note that though we have used a perturbation approach, however, due to the applied limiting procedure the result, Eq. (30), is exact.

The zeros of the hypergeometric function of Eq. (30) have been studied by many authors. A review on this subject is presented in [8]. For a fixed $\delta_0$ there exists infinite sequence of pairs $\{\delta_{11}, \delta_{12}\}$ each of which corresponds to a particular return-resonance of some order $n$, $n \in N$. And for fixed $\delta_0$ and $n$ the pair $\{\delta_{11}, \delta_{12}\}$ is unique. An important property of the zeros is that for any $n$ holds $\delta_{11}(\delta_0, n) = -\delta_{12}(-\delta_0, n)$ so that the two roots, $\delta_{11}$ and $\delta_{12}$ always have different signs. With this observation, we conclude that for fixed finite $\delta_0$ and $n$ the curve in $\{U_{0n}, \delta_1\}$ plane defined by Eq. (26) is an ellipse, which crosses $U_{0n}$-axis at the points $\pm n$ and $\delta_1$-axis at the points $\delta_{11}$ and $\delta_{12}$. The upper half-ellipses for several values of $\delta_0$ for the first resonance $n = 1$ are shown in Fig. 2. These curves are the $\delta_0 = const$ sections of the general surface $U_{01} = U_{01}(\delta_0, \delta_1)$ defining the position of the first return-resonance. This surface is shown in Fig. 3. From the point of view of the generality, the analytical description of this surface for the whole variation range of all involved parameters is the main result of the present paper.

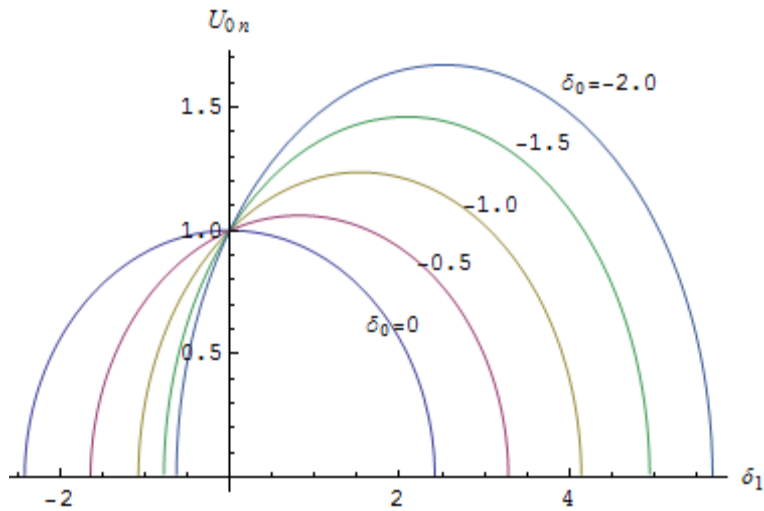

Fig.2. The ellipses defined by Eq. (26) for $n = 1$ and several $\delta_0$.

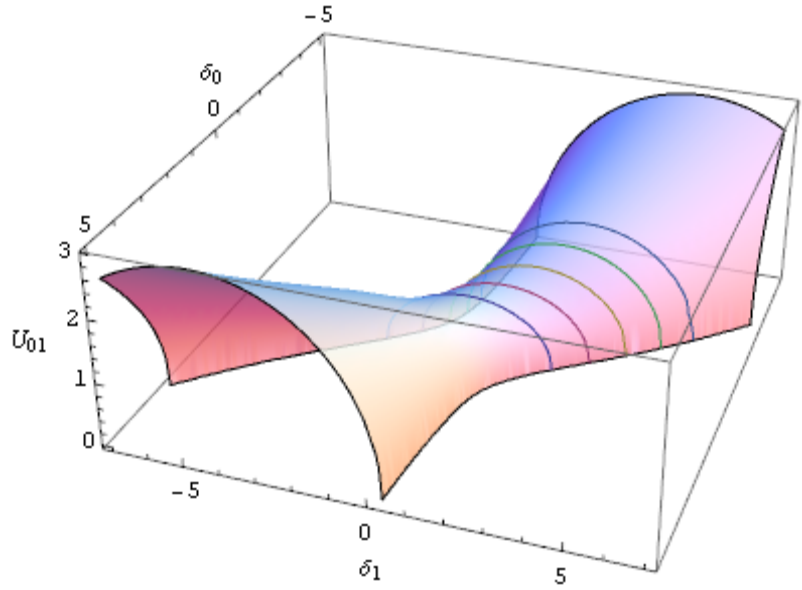

Fig.3. The surface $U_{0n} = U_{0n}(\delta_0, \delta_1)$ defined by Eq. (26) for $n = 1$.

**Acknowledgments**

This research has been conducted within the scope of the International Associated Laboratory (CNRS-France & SCS-Armenia) IRMAS. The research has received funding from the European Union Seventh Framework Programme (FP7/2007-2013) under grant agreement No. 205025 – IPERA. The work has been supported by the Armenian State Committee of Science (SCS Grant No. 11RB-026).

**References**

1. B.W. Shore, *The Theory of Coherent Atomic Excitation* (Wiley, New York, 1990).
2. L. Allen, J.H. Eberly, *Optical Resonance and Two-Level Atoms* (Wiley, New York, 1975).
3. N. Rosen and C. Zener. Phys. Rev., **40**, 502 (1932).
4. A. Ronveaux, *Heun's Differential Equations* (Oxford University Press, London, 1995).
5. I.I. Rabi, Phys. Rev. **51**, 652 (1937).
6. A. Bambini and P.R. Berman, Phys. Rev. A **23**, 2496 (1981).
7. A. Ishkhanyan, R. Sokhoyan, B. Joulakian, and K.-A. Suominen, Optics Commun. **282**, 218 (2009).
8. G.N. Georgiev and M.N. Georgieva-Grosse, J. Telecommun. Inf. Techn. **3**, 112 (2005).